\documentclass[aps,reprint]{revtex4-2}

\usepackage{
amsmath,
amssymb,
graphicx
}
\usepackage{multirow}

\begin{document}

\title{Hybrid Skyrmions in Magnetic Multilayer Thin Films are Half-Integer Hopfions}

\author{W. S. Parker}
\email{wparker4@uoregon.edu}
\author{J. A. Reddinger}
\author{B. J. McMorran}
\affiliation{
Physics Department, University of Oregon
}

\date{\today}

\begin{abstract}
Magnetic skyrmions are chiral spin textures which have attracted intense research for their fundamentally novel physics and potential applications as spintronic information carriers.
The stability which makes them so potentially useful is a result of their underlying non-trivial topology. 
While skyrmions were originally predicted and observed in crystalline materials lacking inversion symmetry, some of the most promising host systems for skyrmions are multilayer thin films, where skyrmions have been stabilized at ambient conditions, which is critical for their use in real world devices. 
The skyrmions found in multilayer thin films have additional three-dimensional structure, with their domain wall helicities twisting through the thickness of the film to create a hybrid skyrmion composed of a Bloch-type core with Néel-type caps of opposite chiralities at the surfaces. 
In this work, we show that this three-dimensional variation creates additional knotted topological structure, providing an explanation for their exceptional stability in ambient conditions. 
We show that hybrid skyrmions can be described as half-integer Hopfions, 
and that their field lines have the knotted structure of the Hopf fibration. 
Furthermore, we show that the topological charge of partially twisted hybrid skyrmions can be related to the domain wall helicity at the surfaces, providing a straightforward way to connect experimental measurements to underlying topology. 

\end{abstract}


\maketitle 

\section{\label{sec:intro}Introduction}

Magnetic skyrmions are chiral spin textures whose rich underlying physics and potential applications in logic and memory devices have drawn the attention of substantial research focus \cite{desautelsRealizationOrderedMagnetic2019a, garaninWritingSkyrmionsMagnetic2018, gobelSkyrmionsReviewPerspectives2021, jiangSkyrmionsMagneticMultilayers2017, zhouMagneticSkyrmionsIntriguing2019}. 
Their non-trivial topology grants them inherent stability, while their localized, particle-like nature may allow them to bypass pinning potentials when driven by a current \cite{buttnerTheoryIsolatedMagnetic2018, fertSkyrmionsTrack2013a}.
However, their use in real-world devices will require stability in ambient conditions (room temperature and with no applied field) and efficient current driven motion, both of which have proven difficult to achieve \cite{nagaosaTopologicalPropertiesDynamics2013, jiangSkyrmionsMagneticMultilayers2017}. 
This is in part because while skyrmions are two-dimensional solitons, the systems that host them are three-dimensional.  
Magnetic skyrmions were first observed in bulk MnSi crystals, first via small angle neutron scattering \cite{muhlbauerSkyrmionLatticeChiral2009} and later with Lorentz transmission electron microscopy \cite{yuRealspaceObservationTwodimensional2010}. 
The non-centrosymmetric crystal structure in these materials leads to a non-collinear Dzyaloshinskii-Moriya Interaction (DMI) term which causes a canting between pairs of spins \cite{muhlbauerSkyrmionLatticeChiral2009, yuRealspaceObservationTwodimensional2010, nagaosaTopologicalPropertiesDynamics2013}. 
The result is a variety of spin textures whose form is determined by the crystalline structure.

Multilayer thin films are one of the few systems capable of hosting stable magnetic skyrmions at ambient conditions (room temperature and no applied field), and as a result they are a promising avenue forward for real-world devices \cite{desautelsRealizationOrderedMagnetic2019a, jiangSkyrmionsMagneticMultilayers2017, montoyaResonantPropertiesDipole2017a, montoyaTailoringMagneticEnergies2017b, montoyaTransportPropertiesDipole2022}. 
In particular, by adjusting the number and thickness of layers in amorphous Fe/Gd multilayers, Desautels et. al. were able to create lattices of skyrmions that are stable across a wide range of temperatures and magnetic fields, including ambient conditions \cite{desautelsRealizationOrderedMagnetic2019a}. 
The skyrmions they form have a hybrid domain wall helicity structure which is able to bring magnetic flux-closures into the material, decreasing its magnetostatic energy, and their exceptional stability is a result of this hybrid structure. 
In these systems, skyrmions are stabilized by the competition between long-range dipole and short-range exchange interactions, and the lack of a symmetry-breaking mechanism allows skyrmions of opposite chiralities to coexist. 
Recent work has shown that the surface-volume stray field interactions in these films have the form of a layer-dependent interfacial DMI \cite{lemeshTwistedDomainWalls2018, liyanageThreedimensionalStructureHybrid2023}, causing the naturally Bloch-type domain walls (DWs) to twist toward Néel-type at the surfaces, with opposite helicities at top and bottom. 
These Néel caps function as flux closure domains, acting to bring stray dipolar fields inside the film and lower the magnetostatic energy. 
The result is a fundamentally three-dimensional structure known as a hybrid skyrmion, which is not seen in conventional DMI-based skyrmion systems. 
This has a significant effect on stability as well as current-driven motion, as both velocity response and Hall angle depend on skyrmion helicity \cite{lemeshTwistedDomainWalls2018}. 
Understanding this additional 3D structure is therefore critical for predicting and tailoring skyrmion properties towards real world applications. 

Here, we show that the DW twisting in hybrid skyrmions creates additional topological structure, contributing to their exceptional stability. 
Specifically, hybrid skyrmions can be modeled as half-integer Hopfions - three-dimensional solitons with Hopf index $\pm1/2$ - and we show how they can be constructed directly from the Hopf fibration underlying integer Hopfions.
Partially twisted DWs lead to a fractional Hopf index between $0$ and $\pm 1/2$, which provides an important metric for experimental comparison, allowing topological charge to be related solely to surface-sensitive measurements. 
The hybrid skyrmions' fractional charge is a result of unsatisfied boundary conditions; while integer Hopfions are localized in three dimensions, hybrid skyrmions are localized in only two, with the geometry of the system constraining the third. 
This non-localization is a general feature of solitons with fractional charge, for example two-dimensional magnetic vortices (merons)\cite{xiaQubitsBasedMerons2022}, Bloch points \cite{tejoBlochPoint3D2021}, and optical vortices \cite{coulletOpticalVortices1989}. 

We use MuMax3 to perform micromagnetic simulations representative of the Fe/Gd multilayer thin films that Desautels et. al. tuned to host skyrmion lattices at room-temperature and remanence, and numerically calculate the Hopf index of the results. 
Our results show good agreement between theoretical expectations, previous observations of this material, and simulated magnetization, and reveal the underlying topology of the complex structure of a hybrid skyrmion. 

\section{\label{sec:background}Background}

Magnetic skyrmions are two-dimensional topological solitons - stable, particle-like configurations of the vector field of magnetization. 
Their non-triviality is captured by a 2D topological index, the skyrmion number $N_{sk} = \frac{1}{4\pi}\int\mathbf{m}\cdot\left[\partial_{x}\mathbf{m}\times\partial_{y}\mathbf{m}\right]dxdy$ where $\mathbf{m}$ is the normalized magnetization vector. 
This index separates two-dimensional vector fields into equivalence classes which cannot be continuously deformed into each other. 
In this way, the topological charge corresponds to physical stability, forbidding smooth transformations between skyrmion states and uniform magnetization. 

The magnetization configuration of a skyrmion can be described as a closed DW loop, separating an out-of-plane magnetic domain from an antiparallel surrounding region. 
The magnetization rotates smoothly from within the domain to without, creating an in-plane DW between the two. 
The orientation of the DW magnetization determines its helicity; Néel-type DWs point normal to the DW itself, while Bloch-type DWs point along the DW. 
In-plane skyrmion analogs, in which the isolated domain and surrounding region lie in-plane, are called bimerons \cite{gobelMagneticBimeronsSkyrmion2019} and have different geometric nuances but equivalent topology. 
In some thin films, skyrmions have a hybrid texture, with Bloch-type DWs at the center of the film and Néel-type DWs at the top and bottom surfaces. 

Hopfions, or Hopf solitons, are a natural way to extend skyrmions into three dimensions. 
The configuration of a Hopfion is that of a skyrmion tube - a skyrmion extended trivially along its central axis - wrapped into a torus. 
Their non-trivial topology is captured by the Hopf index, written in real space as $H=-\frac{1}{(8\pi)^2} \int \mathbf{F}\cdot\mathbf{A} d^3\mathbf{r}$ where $\mathbf{F}$ is the emergent magnetic field and $\mathbf{A}$ its vector potential \cite{whiteheadExpressionHopfInvariant1947}. 
Integer-valued magnetic Hopfions take the form of a bimeron tube wrapped into a closed loop around the $z$-axis, such that the magnetization in the core of the torus is parallel to this axis, while the magnetization outside the torus is antiparallel. 
In this configuration, the core of the torus is an isolated magnetic domain, and the texture as a whole is localized in three dimensions.
These have recently been observed in magnetic systems \cite{kentCreationObservationHopfions2021}, having been studied previously in other contexts, e.g., high energy physics \cite{faddeevStableKnotlikeStructures1997} and superconductivity \cite{babaevHiddenSymmetryKnot2002}.

\begin{figure}
    \centering
    \includegraphics[width=\linewidth]{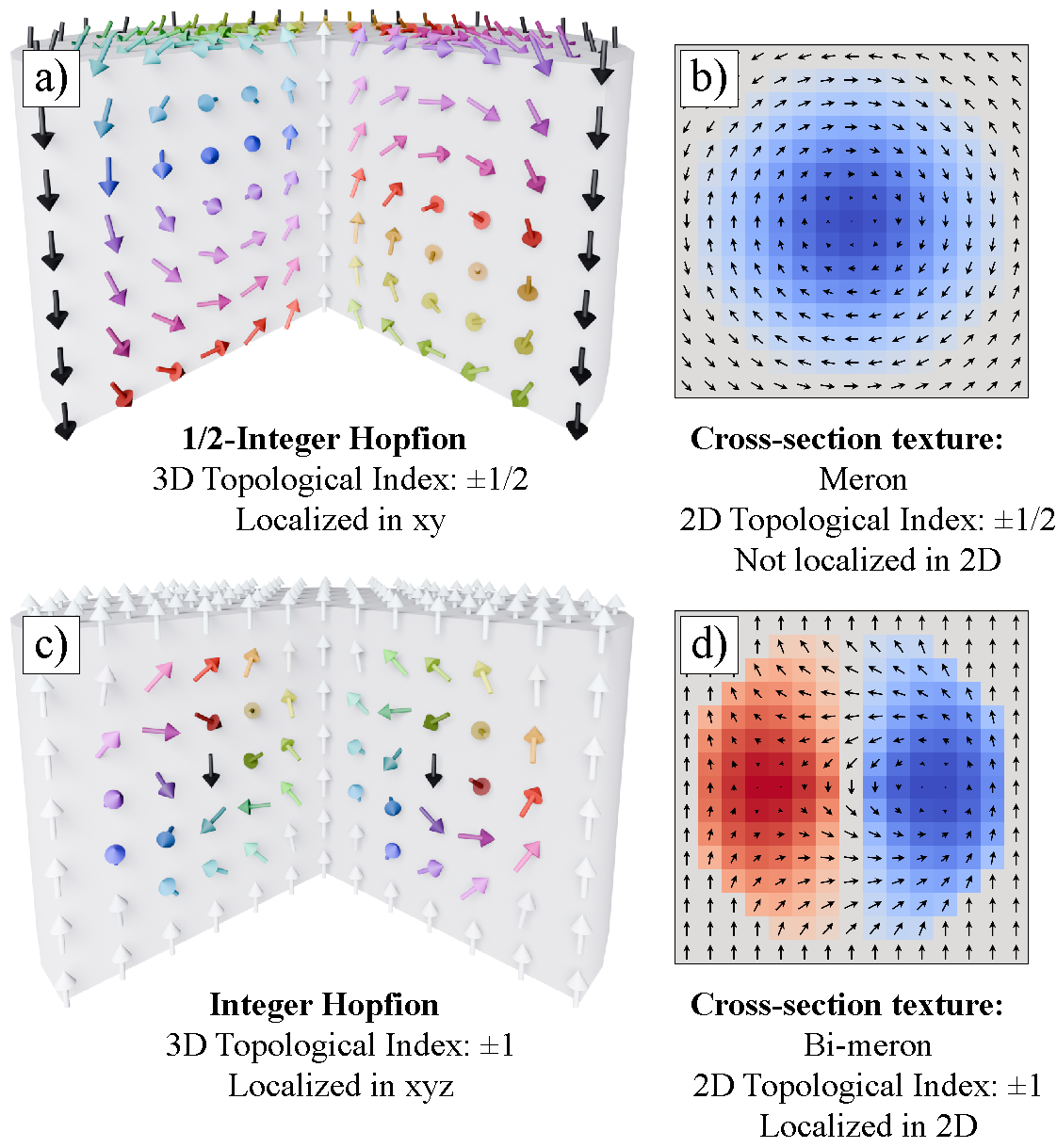}
    \caption{Schematic of a) a hybrid skyrmion, or 1/2-integer Hopfion, and c) an integer Hopfion. The hybrid skyrmion is composed of a b) meron wrapped around the vertical axis, resulting in an isolated magnetic domain along the vertical axis, and is therefore localized only in x and y. The hopfion is composed of a c) bi-meron wrapped around the vertical axis, resulting in a ring-shaped isolated magnetic domain, and is therefore localized in x, y, and z. The meron has 1/2-integer skyrmion number, while the bi-meron (an in-plane skyrmion analogue) has an integer skyrmion number, which explains the 1/2-integer and integer 3D topological charges of the hybrid skyrmion and the Hopfion, respectively.}
    \label{fig:cross-section}
\end{figure}

Here we generalize the Hopfion model by considering a wider class of 2D topological structures (skyrmions) wrapped azimuthally into a torus. 
For a hybrid skyrmion, the magnetic configuration can be achieved by wrapping a half-integer skyrmion (meron) in this way (as opposed to the integer-valued bimeron that results in an integer-valued Hopfion).
This can be seen in the cross-section of its DW, shown in Fig. \ref{fig:cross-section}. 
This has two remarkable effects.
First, the resulting texture has a half-integer Hopf index, since the Hopf index is proportional to this cross-sectional texture's skyrmion number.
Second, the texture can be created from the Hopf fibration by treating the fibers of the Hopf map as field lines of the magnetization; that is, creating a magnetization which is everywhere tangent to the Hopf fibration. 
Physically, the latter point corresponds to bringing flux closure lines inside the material, allowing the magnetization itself to take on some of the skyrmion's dipolar structure. 

\section{\label{sec:hopf_model}Construction of Hybrid Skyrmions from the Hopf Fibration}

\begin{figure}
    \centering
    \includegraphics[width=\linewidth]{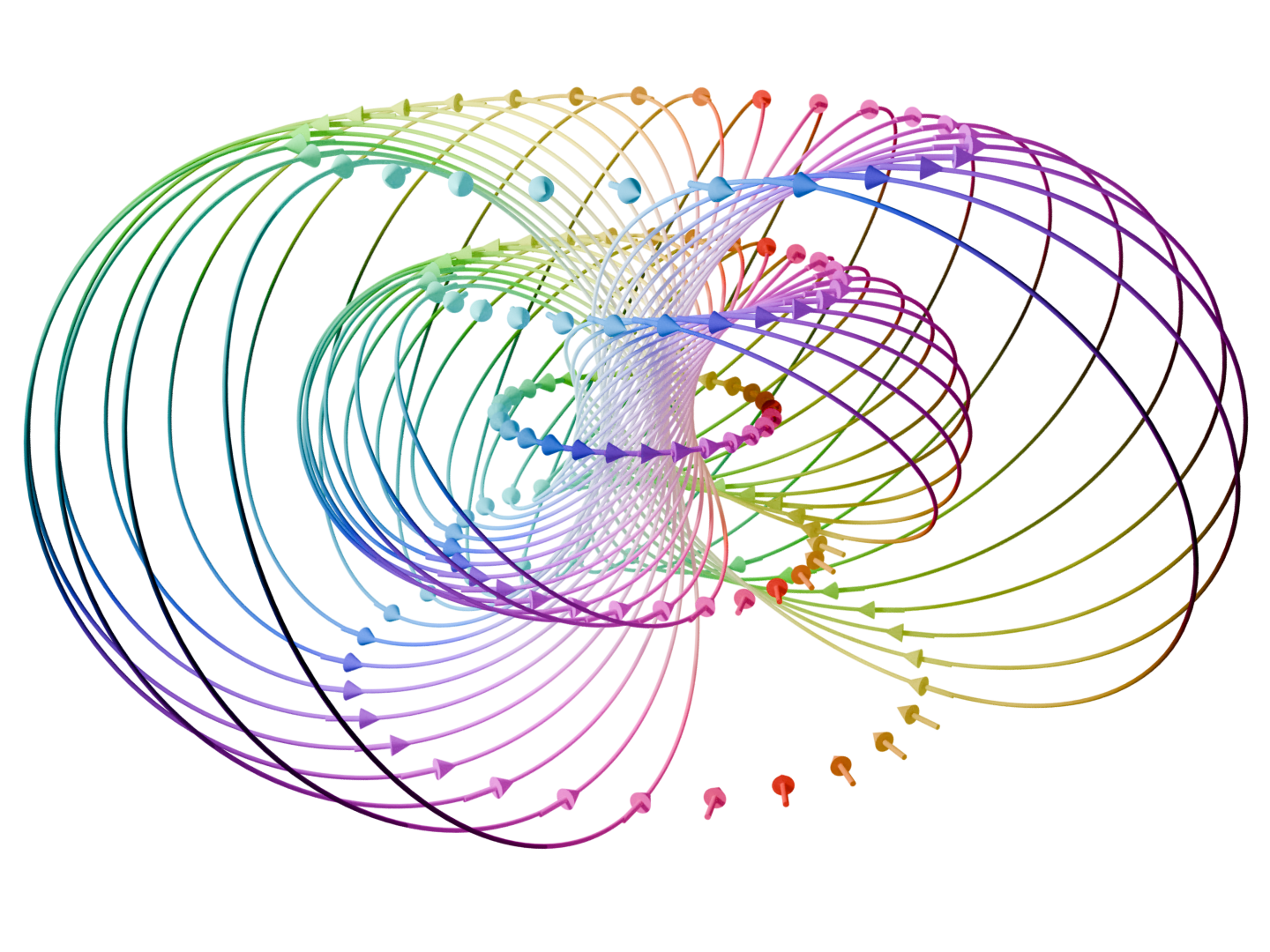}
    \caption{The Hopf fibration in real space. Tangent vectors are plotted at the highest and lowest points of the fibers, where they lie in-plane. These points correspond to the DW center at a given $z$. }
    \label{fig:hopf-fiber-tangents}
\end{figure}

The Hopf fibration fills all of real space $\mathbb{R}^3$ with linked torus knots on a set of nested tori, via stereographic projection of the fibers of the Hopf map from the 3-sphere. 
The vector field of a hybrid skyrmion is formed by taking these torus knots to be its field lines; that is, by constructing a vector field which is at every point in space tangent to the Hopf fiber passing through that point. Fig. \ref{fig:hopf-fiber-tangents} shows these fibers as well as the tangent vectors at their highest and lowest points, which form the centers of the skyrmion's DWs. 
Note the distinction from integer Hopfions: for integer Hopfions, the Hopf fibers define curves along which magnetization is constant, while for hybrid skyrmions the Hopf fibers are everywhere tangent to the magnetization. 
As a result, in hybrid skyrmions the fibers act as magnetic flux lines, and their knotted structure directly implies their resistance to annihilation. 

As shown in Appx. \ref{sec:Appendix1}, the Hopf fiber passing through a point $(x, y, z)$ can be written as a parametric curve

\begin{equation}\label{eq:HopfFiberR3}
    \begin{split}
            \mathbf{S}(t) = 
    \frac{1}{1+R^2+(1-R^2)\cos t - 2 z \sin t} \\
    \times
    \begin{pmatrix}
        2 x\cos t - 2 y\sin t) \\
        2 y\cos t + 2 x\sin t) \\
        2 z \cos t + (1 - R^2) \sin t 
    \end{pmatrix}
    \end{split}
\end{equation}

where $R^2 = x^2 + y^2 + z^2$. 
The hybrid skyrmion magnetization is given by the normalized tangent vectors to this curve

\begin{equation}\label{eq:HIHopfion}
    \begin{aligned}
        \mathbf{m}(x, y, z) &= \frac{\partial_t S(t)}{|\partial_t S(t)|} \Big|_{t=0}
        \\
        &= \frac{1}{1+R^2} 
        \begin{pmatrix}
            2(xz - y) \\ 2(yz + x) \\ 1 + z^2 - x^2 - y^2
        \end{pmatrix}.
    \end{aligned}
\end{equation}

Since this vector field is cylindrically symmetric, it is useful to express it in cylindrical coordinates $(\rho, \phi, z)$

\begin{equation}\label{eq:HIHopfionCylindrical}
    \mathbf{m}(\rho, \phi, z) = 
    \frac{1}{1+\rho^2 + z^2}
    \begin{pmatrix}
        2 \rho (z\cos\phi - \sin\phi) \\ 2\rho (z\sin\phi + \cos\phi) \\ 1+z^2 - \rho^2
    \end{pmatrix}. 
\end{equation}

Eq. (\ref{eq:HIHopfionCylindrical}) is a unit vector field in $\mathbb{R}^3$ which at every point is tangent to the Hopf fiber passing through it.
A similar structure has been found as a solution to a certain class of equations in magnetohydrodynamics \cite{kamchatnovTopologicalSolitonMagnetohydrodynamics2004}, but in that case the tangent was taken prior to stereographic projection, yielding a vector field whose magnitude falls to zero far from the origin and is therefore truly localized. 

\subsection{\label{sec:behavior}Properties and Behavior}

At $z=0$, Eq. (\ref{eq:HIHopfion}) has the form of a Bloch-type skyrmion in the $xy$-plane. 
The magnetization rotates from $\mathbf{m}=+\hat{\mathbf{z}}$ at the origin to $\mathbf{m}=-\hat{\mathbf{z}}$ far from the origin, rotating around $\hat{\rho}$ such that all in-plane components are azimuthal. 
In this plane, the maximum in-plane magnetization is found at $\rho=1$. 

\begin{equation}
    \label{eq:z0plane}
    \mathbf{m}(x, y, 0) = \frac{1}{1+x^2+y^2}
    \begin{pmatrix}
        -2y \\ 2x \\ 1-x^2-y^2
    \end{pmatrix}
\end{equation}

Away from $z=0$, the polarization of the skyrmion - that is, the limits as $\rho\to0,\infty$ - remains the same but the in-plane components rotate, becoming more radial farther from $z=0$. 
As $z$ increases, the in-plane components twist radially outward, creating a Néel-type skyrmion, while as $z$ decreases, they twist radially inward, creating a Néel-type skyrmion of the opposite helicity. 
These helicities match the flux closure curves created by the dipole field of the skyrmion's core and surrounding region and are known as Néel caps. 

The helicity of the skyrmion can be expressed exactly as a function of $z$. 
To do so, we define helicity $\alpha$ as the angle between the in-plane component of magnetization $\mathbf{m}_{\perp} = m_x\hat{\mathbf{x}} + m_y\hat{\mathbf{y}}$ and the in-plane gradient of $m_z$, $\nabla_{\perp}m_z=\partial_x m_z \hat{\mathbf{x}} + \partial_y m_z \hat{\mathbf{y}}$ \cite{chessDeterminationDomainWall2017}. 
With this definition, pure Néel-type skyrmions have helicity either $0$ or $\pi$, while pure Bloch-type skyrmions have helicity $\pi/2$ or $3\pi/2$. 
Calculated explicitly from Eq. (\ref{eq:HIHopfion}), $\alpha$ has the form

\begin{equation}\label{eq:alpha}
    \alpha(z) = \cos^{-1}\left(\frac{-z}{\sqrt{1+z^2}}\right). 
\end{equation}

which varies from $0$ to $\pi$ as $z$ varies from $-\infty$ to $\infty$. 

 In contrast to simulated and experimentally observed hybrid skyrmions, the radius of the Hopf-derived hybrid skyrmion $\rho_{DW}$ - the radius at which the in-plane component is maximized - expands as $|z|$ increases, varying as

\begin{equation}\label{eq:rhoDW}
\rho_{DW}=\sqrt{1+z^2}.
\end{equation}

Micromagnetic simulations show that hybrid skyrmions' DWs are typically barrel-shaped \cite{lemeshTwistedDomainWalls2018}.
As this qualitative difference can be rectified by a smooth coordinate rescaling, it has no impact on the topology of the structure, which is instead determined by the DW helicity.

Lastly, we can calculate the curves along which $\mathbf{m}$ is constant - that is, the fibers of the hybrid skyrmion. 
These are given implicitly by

\begin{equation}
    \label{eq:HIHopfionPreimages}
    \begin{aligned}
        x &= x_0 - y_0 z \\
        y &= y_0 + x_0 z
    \end{aligned}. 
\end{equation}

For any fixed $\rho_0 = \sqrt{x_0^2 + y_0^2}$, these lines, as shown in Fig. \ref{fig:constant-magnetizations}, are the generators of the hyperboloid $\rho=\rho_0\sqrt{1+z^2}$; for example, the hyperboloid given by the skyrmion radius in Eq. (\ref{eq:rhoDW}) when $\rho_0=1$. 
Fig. \ref{fig:constant-magnetizations}(a) shows these curves for the DWs ($\rho_0=1$), while Fig. \ref{fig:constant-magnetizations}(b) shows the same, calculated numerically, for a simulated hybrid skyrmion. 

This last result sheds some light on the hybrid skyrmion's non-integer Hopf index. 
For integer Hopfions, the fibers form closed loops, with a linking number equal to the Hopf index; however, the fibers of Eq. (\ref{eq:HIHopfionPreimages}) extend to infinity, never closing on each other. 
Since stereographic projection maps all infinities in $\mathbb{R}^3$ to a single point in $\mathbb{S}^3$, and the Hopf-derived hybrid skyrmion is not homogeneous at infinity, the hybrid skyrmion is not only many-to-one, but also one-to-many. 
In fact, the pole of $\mathbb{S}^3$ corresponding to infinity in $\mathbb{R}^3$ itself covers all of $\mathbb{S}^2$. 
In other words, a hybrid skyrmion's magnetization approaches a different value far from the origin depending on which direction away from the origin you travel; this incongruity expresses the non-localization of a hybrid skyrmion in 3D. 
In reality, the hybrid skyrmion is localized in $z$ by the finite thickness of the host system. 

\subsection{\label{sec:HopfIndex}Hopf Index}

The Hopf index of a vector field $\mathbf{m}(x,y,z)$ can be calculated explicitly using its real-space expression \cite{whiteheadExpressionHopfInvariant1947} 

\begin{equation}\label{eq:HopfIndexRaw}
    H = -\frac{1}{(8\pi)^2} \int \mathbf{F}\cdot\mathbf{A} d^3\mathbf{r}
\end{equation}

where $F_i = \epsilon_{ijk} \mathbf{m}\cdot(\partial_j\mathbf{m}\times\partial_k\mathbf{m})$ and $\nabla\times\mathbf{A}=\mathbf{F}$ are additionally defined based on $\mathbf{m}$. 
$\mathbf{F}$ can be seen as the emergent magnetic field, and $\mathbf{A}$ its vector potential \cite{kentCreationObservationHopfions2021}. 

We can parameterize the magnetization in a natural way by azimuthal and polar angles $\Phi$ and $\Theta$, respectively: $\mathbf{m}=(\sin\Theta\cos\Phi, \sin\Theta\sin\Phi, \cos\Theta)$.
When $\Theta$ is independent of $\phi$ and $\Phi$ has the form $\Phi = Q\phi + h(\rho, z)$ (for some integer $Q$ and some function $h(\rho, z)$ which is independent of $\phi$), the Hopf index can be expressed in the form \cite{Gladikowski_1997}

\begin{equation}
    \begin{aligned}
            H &= \frac{Q}{4\pi} \int_{-\infty}^{\infty}dz\int_{0}^{\infty}dr \sin\Theta\left(
            \frac{\partial\Theta}{\partial\rho}\frac{\partial h}{\partial z} - \frac{\partial\Theta}{\partial z}\frac{\partial h}{\partial \rho}
            \right)
            \\
            &= Q \times N_{sk}(\rho, z)
    \end{aligned}
\end{equation}

where $N_{sk}(\rho, z)$ is the skyrmion number of the toroidally wrapped texture, and $Q$ is the winding number around the $z$ axis. 
Here, the Hopf index takes the explicit form of the product of these two quantities. 
As noted above, for hybrid skyrmions the toroidally wrapped texture is a meron, with half-integer charge, while for integer Hopfions the toroidally wrapped texture is a bimeron, with integer charge. 
$\Theta$ and $\Phi$ can be calculated directly from our Hopf-derived model of the hybrid skyrmion, Eq. (\ref{eq:HIHopfionCylindrical}), and $Q=\partial_{\phi}\Phi=1$, giving an analytical solution for $H$: 

\begin{equation}\label{eq:HopfIndex}
    \begin{aligned}
            H &= \frac{1}{4\pi} \int_{-\infty}^{\infty}dz\int_{0}^{\infty}dr \left(
    -\frac{4\rho}{(1+z^2+\rho^2)^2}\right) 
    \\
    &= -\frac{1}{2}
    \end{aligned}
\end{equation}

Note that in the construction of $\mathbf{m}$, Eq. (\ref{eq:HIHopfion}), we had a choice of two directions for the Hopf fibers' tangent vectors; making the opposite choice results in $H=+1/2$. 
Like the skyrmion index, the sign of the Hopf index depends on the polarity of the spin texture. 

\subsection{\label{sec:helicityDependentHopfIndex}Hopf index of partially twisted skyrmions}

In real systems, the Néel caps arise from a layer dependent interaction competing with naturally Bloch-type domain walls. 
As a result, the surfaces of the material are not necessarily completely Néel-type, but instead retain some azimuthal component. 
In addition, when the system contains inherent DMI, the Bloch core may be displaced above or below the center of the film \cite{lemeshTwistedDomainWalls2018}. 
Here, we consider the Hopf index of such a partially twisted, possibly asymmetric, hybrid skyrmion by recalculating the Hopf index with restricted limits of integration. 

If we restrict the integration w.r.t. $z$ in Eq. (\ref{eq:HopfIndex}) to a lower limit $z_l$ and an upper limit $z_u$, we find that 

\begin{equation}\label{eq:zlimHIndex}
    H = \frac{\tan^{-1}(z_l)-\tan^{-1}(z_u)}{2\pi}. 
\end{equation}

We can rewrite this in terms of the helicity at the surfaces by solving for $z$ in Eq. (\ref{eq:alpha}), and substituting the result into Eq. (\ref{eq:zlimHIndex}), giving

\begin{equation}\label{eq:alphaHIndex}
    H = \frac{\tan^{-1}(\cot(\alpha_u))-\tan^{-1}(\cot(\alpha_l))}{2\pi},
\end{equation}

where $\alpha_l$, $\alpha_u$ are the helicities at $z_l$, $z_u$, respectively. 
In the symmetric case $z_l=-z_u$, the Hopf index is given in terms of helicity by 

\begin{equation}\label{eq:alphaHIndexSymmetric}
    H=\frac{\tan^{-1}(\cot(\alpha_u))}{\pi}
\end{equation}

This means the Hopf index varies continuously from $0$ when the skyrmion is uniform throughout the thickness, to $\pm1/2$ when the skyrmion twists to fully Néel-type at the surfaces. 
For a texture where the Bloch walls twist only partially toward Néel-type at the surface, the Hopf index is between $0$ and $\pm1/2$.
Furthermore, in this case, the Hopf index depends solely on the helicity at the surface. 
This is an important result for the experimental characterization of hybrid skyrmions, as it means the 3D topological charge can be understood using only surface-sensitive magnetic imaging - provided the texture is known to have this hybrid structure - without the need to quantitatively measure the full 3D magnetization.

\section{\label{sec:experimental}Micromagnetic Simulations}

To assess the validity of these results in the context of a well-studied system, we numerically solved the Landau-Lifshitz-Gilbert (LLG) equation using MuMax3 \cite{vansteenkisteDesignVerificationMuMax32014}. 
The material parameters were chosen to match those obtained experimentally by Montoya et. al in \cite{montoyaTailoringMagneticEnergies2017b}, which are representative of an Fe/Gd multilayer thin film, and temperature was set to 300K. 
A $1.28$µm$\times1.28$µm$\times80$nm slab was initialized with random magnetization. 
A saturating perpendicular field of $-330$mT was initialized, then decreased in increments of 5mT, with the LLG equation evolved for 15ns at each field step. 
This procedure generates a stable mixture of worm domains and skyrmions at remanence. 

Next, we simulated a field sweep from remanence back to negative saturation, allowing us to predict the behavior of the Néel caps and associated evolution of the Hopf index of a hybrid skyrmion as they varied with applied field. 
The applied field was decreased in steps of 5mT from remanence, and the LLG equation evolved for 15ns at each field step, until all magnetic domains were annihilated at $-325$mT. 

Additionally, the magnetization after the $-250$mT field step was used as a seed for a high-resolution simulation using the \verb|relax()| function of MuMax3, which disables precession in the LLG equation to instead attempt to find the energetic minimum \cite{vansteenkisteDesignVerificationMuMax32014}. 
This method produces a less noisy result, while preserving the structure. 
At $-250$mT, all domains have shrunken to isolated skyrmions, so this additional simulation allows for a closer examination of the structure of an individual hybrid skyrmion.

\section{\label{sec:results}Results}

\begin{figure}
    \centering
    \includegraphics[width=\linewidth]{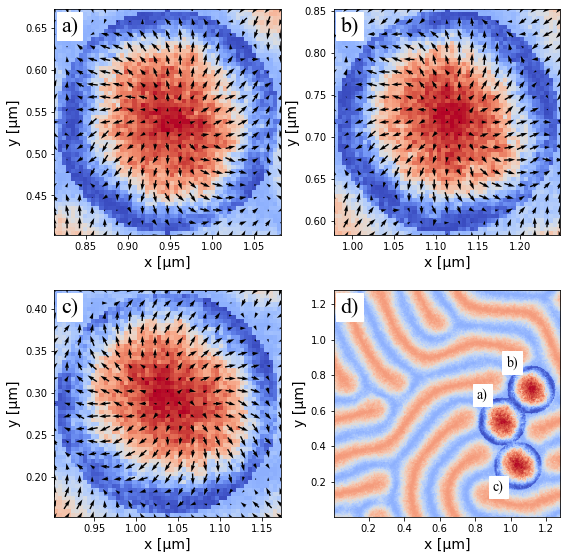}
    \caption{Simulated surface magnetization of three hybrid skyrmions stabilized at remanence. This surface helicity represents the Néel cap, which retains some of the azimuthal component at the center of the film, which is left-handed in a) and b), and right-handed in c). }
    \label{fig:three-skyrmions}
\end{figure}

Fig. \ref{fig:three-skyrmions} shows the surface magnetization of the simulation result at remanence. 
As expected, the DWs are nearly Néel-type, but not fully, competing energetically with the azimuthal component at the center of the film. 
At remanence, we see primarily worm domains, with three skyrmions - two left-handed and one right-handed. 
To quantify the twist of each skyrmion, the surface helicity was averaged for a circular region encompassing the skyrmion, weighted by the magnitude of the in-plane component of magnetization. 
Eq. (\ref{eq:alphaHIndexSymmetric}) was then used to predict the Hopf index we should expect based on the surface helicity. 
The Hopf index was also calculated numerically from the full vector magnetization using the procedure described in Appx. \ref{sec:Appendix3}, and shows a good match to the value predicted by Eq. (\ref{eq:alphaHIndexSymmetric}). 
Table \ref{table:helicities_and_hopf_indexes} summarizes these results. 

\begin{table}
\centering
\begin{tabular}{ |c||c|c|c| } 
 \hline
  & A & B & C \\
 \hline
 Avg. Surface Helicity & $0.8804\pi$ & $0.8843\pi$  & $-0.8757\pi$\\ 
 Helicity-based HI (Eq. (\ref{eq:alphaHIndexSymmetric})) & $-0.3804$ & $-0.3843$ & $0.3757$\\ 
 Numerically-calculated HI & $-0.3909$ & $-0.3831$ & $0.3495$\\
 \hline
\end{tabular}
\caption{The average surface helicity, surface-helicity based Hopf index (calculated using Eq. (\ref{eq:alphaHIndexSymmetric})), and numerically calculated Hopf index of the simulated skyrmions shown in Fig. \ref{fig:three-skyrmions}. }
\label{table:helicities_and_hopf_indexes}
\end{table}

\begin{figure}
    \centering
    \includegraphics[width=\linewidth]{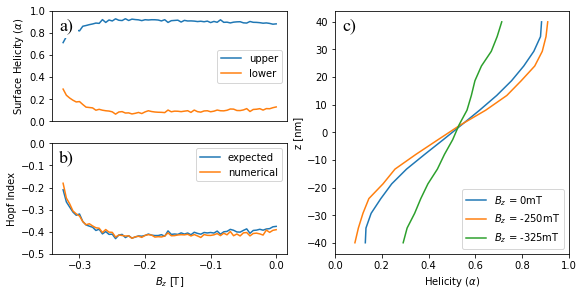}
    \caption{Evolution of (a) surface helicity and (b) Hopf index of the simulated skyrmion as $B_z$ is decreased from remanence to $-325$mT. (c) shows the average helicity through the thickness of the skyrmion for three applied field values. }
    \label{fig:helicity-plots}
\end{figure}

Next, a field sweep from remanence to saturation was simulated, using the magnetization shown in Fig. \ref{fig:three-skyrmions}(d) as a seed. 
The field was decreased in steps of 5mT, and the LLG equation evolved for 15ns at each step, until all domains were annihilated at a field value of -325mT. 
At each step, the skyrmion's surface helicities and numerical Hopf index were calculated, with its center tracked by the weighted average of its core magnetization. 
Fig. \ref{fig:helicity-plots}(a) shows the calculated average surface helicities at the upper and lower surfaces of the film, and Fig. \ref{fig:helicity-plots}(b) shows the Hopf index expected from these surface helicities, along with the explicitly calculated Hopf index. 
Fig. \ref{fig:helicity-plots}(c) shows the $z$-dependent average helicity for three different field values: remanence, the value at which surface helicity was closest to Néel-type, and the last field step before annihilation. 

\begin{figure}
    \centering
    \includegraphics[width=\linewidth]{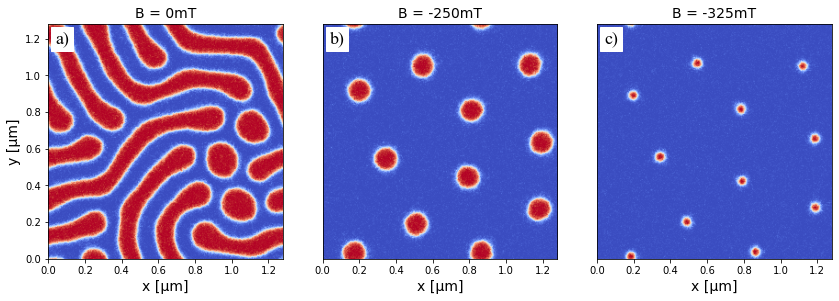}
    \caption{Z-component of magnetization at the center of the film at a) remanence, b) -250mT, and c) -325mT. }
    \label{fig:three-field-values}
\end{figure}

Notably, the Néel caps are most Néel-type at around -220mT, rather than near remanence or saturation. 
At remanence, the DWs of neighboring magnetic domains are pressed together (Fig. \ref{fig:three-field-values}(a)), and their Néel caps antiparallel, making it energetically unfavorable for the DWs to twist fully Néel-type. 
As the magnitude of applied field is increased, skyrmions become more isolated (Fig. \ref{fig:three-field-values}(b)), allowing their Néel caps to twist more fully. 
However, the skyrmions also shrink with applied field, until the DWs on opposite sides of an individual skyrmion begin to interact (Fig. \ref{fig:three-field-values}(c)), again creating a preference against fully twisted Néel caps. 

\begin{figure}
    \centering
    \includegraphics[width=\linewidth]{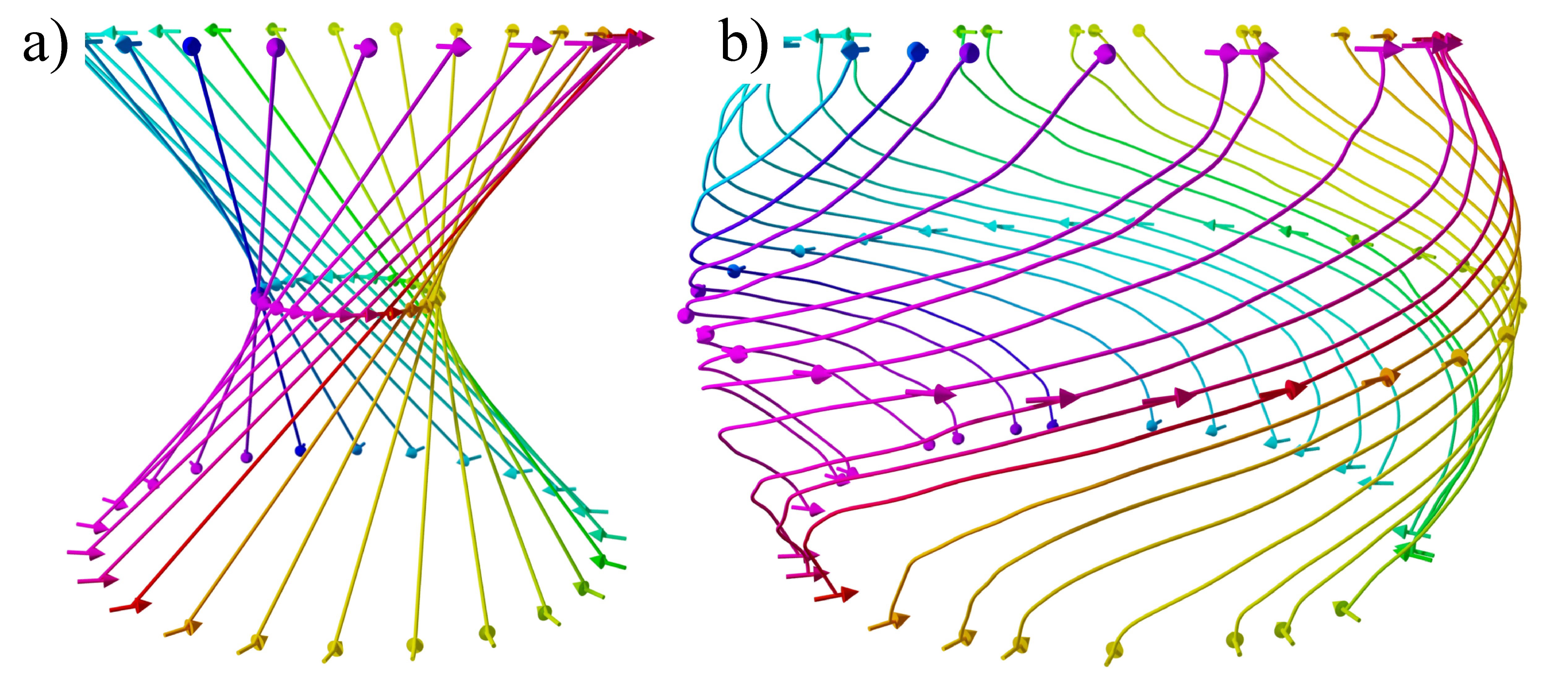}
    \caption{Curves of constant magnetization for (a) the skyrmion constructed from the Hopf fibration, Eq. (\ref{eq:HIHopfion}), and (b) the simulated isolated hybrid skyrmion, starting from points at the center of the DW wall. Note how the curves wrap almost but not quite 180º around the $z$-axis; this is equivalent to the nearly 180º twist in helicity from top to bottom. The $z$-dependence of this wrapping matches the $z$-dependence of its helicity (Fig. \ref{fig:helicity-plots}(c), $B_z$=-0.250T), varying fastest at the center and slowing near the surfaces. }
    \label{fig:constant-magnetizations}
\end{figure}

Finally, the $-250$mT magnetization, after relaxation to a minimum energy state, was used to numerically calculate the curves of constant magnetization for a hybrid skyrmion. 
Fig. \ref{fig:constant-magnetizations}(a) shows these curves for the half-integer Hopfion constructed from the Hopf fibration, Eq. (\ref{eq:HIHopfion}), and Fig. \ref{fig:constant-magnetizations}(b) shows the numerically calculated results for a simulated hybrid skyrmion. 
The simulated hybrid skyrmion's DWs are barrel-shaped rather than hyperboloid, but can be created from the former's via a smooth rescaling of radial coordinates. 
This barrel shape is also predicted by \cite{lemeshTwistedDomainWalls2018}. 
The salient feature is the nearly 180º rotation from top to bottom, which corresponds to the varying helicity.

\section{\label{sec:conclusion}Conclusion}

As one of the few systems capable of hosting stable magnetic skyrmions at ambient conditions, multilayer thin films are a promising avenue forward for real-world devices. 
The skyrmions they host have a hybrid helicity structure, resulting in exceptional stability, with the Néel caps bringing the skyrmions' magnetic flux closures into the material and decreasing their magnetostatic energy. 
In this paper we showed that this additional twisting creates a knotted topological structure.
In particular, hybrid skyrmions have a half-integer Hopf index, adding another non-trivial topology to the plethora of topological spin textures observed in magnetic systems. 
We also showed that the structure of a hybrid skyrmion can be constructed directly by creating a vector field tangent to the fibers of the Hopf fibration. 
This has the physical interpretation of tying magnetic field lines into linked torus knots, creating a barrier against the annihilation of the structures. 
We used micromagnetic simulations to tie these results to a well known and highly promising system, Fe/Gd multilayers, in order to understand the underlying topology first at remanence as well as in the presence of an external field. 
As new chiral spin textures continue to be observed across a range of systems, it is vital to understand the three-dimensional structure and topology of each new texture in order to predict and understand its dynamics, properties, and stability.

\begin{acknowledgments}

We thank Jordan Chess and Tzula Propp for their work creating an initial tunable analytical model of hybrid skyrmions and applying the Hopf index to it.
This work benefited from access to the University of Oregon high performance computing cluster, Talapas.
This material is based upon work supported by the National Science Foundation under Grant No. 2105400. 
\end{acknowledgments}

\appendix

\section{\label{sec:Appendix1}Detailed Construction From Hopf Fibration}

The Hopf fibration is a set of curves on the 3-sphere $\mathbb{S}^3$ defined by the fibers of the Hopf map - the fibers of a function being all the elements of its domain that map to the same point. 
For example, the set $\{-1, 1\}$ can be considered one fiber of the function $f(x)=x^2$, as $f(-1)=f(1)=1$. 
The Hopf map is the many-to-one continuous function from the 3-sphere to the ordinary 2-sphere $h: \mathbb{S}^3 \to \mathbb{S}^2$ defined by

\begin{equation}
    h(x_1, y_1, x_2, y_2) = 
    \begin{pmatrix}
        2 (x_1 x_2 + y_1 y_2) \\
        2 (y_1 x_2 - x_1 y_2) \\
        x_1^2 + y_1^2 - x_2^2 - y_2^2
    \end{pmatrix},
\end{equation}

whose fibers form circles on the 3-sphere. 
It can be directly verified that if two points $(x_1, y_1, x_2, y_2)$ and $(x_1', y_1', x_2', y_2')$ map to the same point under the Hopf map - that is, $h(x_1, y_1, x_2, y_2) = h(x_1', y_1', x_2', y_2')$ - then the two points can be related as follows

\begin{equation}
    \begin{pmatrix}
        x_1'\\y_1'\\x_2'\\y_2'
    \end{pmatrix}
    = 
    \begin{pmatrix}
        \cos t & - \sin t & 0 & 0 \\
        \sin t & \cos t & 0 & 0 \\
        0 & 0 & \cos t & -\sin t \\
        0 & 0 & \sin t & \cos t
    \end{pmatrix}
    \begin{pmatrix}
        x_1\\y_1\\x_2\\y_2
    \end{pmatrix},
\end{equation}

for any $t\in[0,2\pi)$, defining a circle in $\mathbb{S}^3$. 
We can use this relation to write the Hopf fiber passing through an initial point $(x_1, y_1, x_2, y_2)\in\mathbb{S}^3$ as a parametric curve

\begin{equation}\label{eq:parametricS3}
    \mathbf{s}(t) = 
    \begin{pmatrix}
        x_1 \cos t - y_1 \sin t \\
        x_1 \sin t + y_1 \cos t \\
        x_2 \cos t - y_2 \sin t \\
        x_2 \sin t + y_2 \cos t 
    \end{pmatrix}.
\end{equation}

We can relate this fiber to 3D space $\mathbb{R}^3$ via the stereographic projection, given by 

\begin{equation}
    \begin{pmatrix}
        S_1\\S_2\\S_3
    \end{pmatrix}
    =
    \frac{1}{1-s_4}
    \begin{pmatrix}
        s_1\\s_2\\s_3
    \end{pmatrix}. 
\end{equation}

Applying this to Eq. (\ref{eq:parametricS3}) gives an equation for the Hopf fiber in $\mathbb{R}^3$

\begin{equation}\label{eq:parametricR3}
\begin{aligned}
    \mathbf{S}(t) = \frac{1}{1-x_2 \sin t - y_2 \cos t} \\
    \times
    \begin{pmatrix}
        x_1 \cos t - y_1 \sin t \\
        x_1 \sin t + y_1 \cos t \\
        x_2 \cos t - y_2 \sin t
    \end{pmatrix}.
\end{aligned}
\end{equation}

However, this is expressed in terms of an initial point $(x_1, y_1, x_2, y_2) \in \mathbb{S}^3$. 
We can relate this initial point to the initial point in $\mathbb{R}^3$ via the inverse stereographic projection, given by 

\begin{equation}\label{eq:InverseSP}
    \begin{pmatrix}
        x_1\\y_1\\x_2\\y_2
    \end{pmatrix}
    =
    \frac{1}{1+R^2}
    \begin{pmatrix}
        2x\\2y\\2z\\R^2 - 1
    \end{pmatrix}
\end{equation}

where $R^2=x^2 + y^2 + z^2$. 
Substitution of Eq. (\ref{eq:InverseSP}) into Eq. (\ref{eq:parametricR3}) gives the result Eq. (\ref{eq:HopfFiberR3}), which is a parametrization of the Hopf fiber which passes through an initial point $(x, y, z)$. 

\section{\label{sec:Appendix3}Numerical calculation of the Hopf index}

To numerically calculate the Hopf index, it is necessary to evaluate Eq. (\ref{eq:HopfIndexRaw}). 
$\mathbf{F}$ contains only spatial derivatives of $\mathbf{m}$, and is therefore straightforward to calculate numerically.
$\mathbf{A}$ is defined implicitly via $\nabla\times\mathbf{A}=\mathbf{F}$ and is therefore less trivial. 
To calculate $\mathbf{A}$, we choose the Coulomb gauge $\nabla\cdot\mathbf{A}=0$, so that we can use the identity

\begin{equation}
        \nabla\times(\nabla\times\mathbf{A}) = \nabla(\nabla\cdot\mathbf{A}) - \nabla^2\mathbf{A}
\end{equation}

which, substituting the implicit defintion of $\mathbf{A}$ on the LHS and the Coulomb gauge condition on the RHS, reduces to

\begin{equation}
        \nabla\times\mathbf{F} = -\nabla^2\mathbf{A}.
\end{equation}

This is Poisson's equation, and it can be numerically solved with a variety of existing techniques - for example, by using Fourier transforms to replace differentiation with multiplication. 

\bibliography{biblio}

\end{document}